\input harvmac.tex
%\input epsf.tex
%
%\draft
%\newcount\figno
\figno=0
\def\fig#1#2#3{
\par\begingroup\parindent=0pt\leftskip=1cm\rightskip=1cm\parindent=0pt
\baselineskip=11pt
\global\advance\figno by 1
\midinsert
\epsfxsize=#3
\centerline{\epsfbox{#2}}
\vskip 12pt
{\bf Fig. \the\figno:} #1\par
\endinsert\endgroup\par
}
\def\figlabel#1{\xdef#1{\the\figno}}
\def\encadremath#1{\vbox{\hrule\hbox{\vrule\kern8pt\vbox{\kern8pt
\hbox{$\displaystyle #1$}\kern8pt}
\kern8pt\vrule}\hrule}}

\overfullrule=0pt

%macros
%
\def\cqg#1#2#3{{\it Class. Quantum Grav.} {\bf #1} (#2) #3}
\def\np#1#2#3{{\it Nucl. Phys.} {\bf B#1} (#2) #3}
\def\pl#1#2#3{{\it Phys. Lett. }{\bf B#1} (#2) #3}

\def\physrev#1#2#3{{\it Phys. Rev.} {\bf D#1} (#2) #3}
\def\ap#1#2#3{{\it Ann. Phys.} {\bf #1} (#2) #3}

\def\cmp#1#2#3{{\it Comm. Math. Phys.} {\bf #1} (#2) #3}
\def\mpl#1#2#3{{\it Mod. Phys. Lett. }{\bf #1} (#2) #3}
\def\ijmp#1#2#3{{\it Int. J. Mod. Phys.} {\bf #1} (#2) #3}

\font\zfont = cmss10 %scaled \magstep1

\def\bigone{\hbox{1\kern -.23em {\rm l}}}
\def\ZZ{\hbox{\zfont Z\kern-.4emZ}}

\def\a{\alpha}
\def\IR{\relax{\rm I\kern-.18em R}}

\def\g{\gamma}
\def\d{\delta}
\def\e{\epsilon}

\def\l{\lambda}
\def\m{\mu}
\def\n{\nu}

\def\r{\rho}

\def\s{\sigma}

\def\G{\Gamma}

\def\O{\Omega}

\Title{
{\vbox{
\rightline{\hbox{hepth/0210131}}
\rightline{\hbox{UMD-PP-03-015}}
}}}
{\vbox{\hbox{\centerline{A Note on Flux Induced Superpotentials
in String Theory}}}}
\smallskip
\centerline{Melanie Becker\footnote{$^1$} {\tt melanieb@physics.umd.edu} and
Drago\c{s}
Constantin\footnote{$^2$}{\tt dragos@physics.umd.edu}} \centerline{\it
Department of  Physics,
University of Maryland} \centerline{\it College Park, MD 20742-4111}

\bigskip
\bigskip
\baselineskip 18pt
\bigskip
\noindent
Non-vanishing fluxes in ${\cal M}$-theory and string theory
compactifications induce a superpotential in the lower dimensional theory.
Gukov has conjectured the explicit form of this superpotential.
We check this conjecture for the
heterotic string compactified
on a Calabi-Yau three-fold as well as for warped ${\cal M}$-theory
compactifications on Spin(7) holonomy manifolds, by performing a
Kaluza-Klein reduction.

\Date{October, 2002}

\newsec{Introduction}
Compactifications of string theory and ${\cal M}$-theory with
non-vanishing expectation values for tensor fields play a very
special role, when trying to find a realistic string theory model
that could describe our four-dimensional world. Especially
interesting are the so called warped compactifications. Such
compactifications were first discovered for the heterotic string
in \ref\str{A.~Strominger, ``Superstrings with Torsion'', \np
{274} {1986} {253}.} and \ref\wsd{B. de Wit, D.~J.~Smit and
N.~D.~Hari Dass, ``Residual Supersymmetry of Compactified $D=10$
Supergravity'', \np {283} {1987} {165}.} and were later
generalized to warped compactifications of ${\cal M}$-theory and
${\cal F}$-theory in \ref\bb{K.~Becker and M.~Becker, ``${\cal
M}$-Theory on Eight Manifolds'', \np {477} {1996} {155},
hep-th/9605053.}, \ref\svw{S.~Sethi, C.~Vafa and E.~Witten,
``Constraints on Low Dimensional String Compactifications'', \np
{480} {1996} {213}, hep-th/9606122.}, \ref\drs{K.~Dasgupta,
G.~Rajesh and S.~Sethi, ``M-Theory, Orientifolds and G-Flux'',
JHEP 9908:023, 1999, hep-th/9908088.}. In these compactifications
tensor fields acquire non-vanishing expectation values, while
leaving supersymmetry unbroken. The restrictions imposed by
supersymmetry on the fluxes lead to constraints on the moduli
fields of the theory, so that most of these fields will be
stabilized. This has important phenomenological implications and
will lead us one step closer to determining the coupling
constants of the standard model out of string theory. The
constraints imposed by supersymmetry for compactifications of
${\cal M}$-theory on a Calabi-Yau
four-fold\foot{Compactifications of Type II theories on Calabi-Yau
four-folds were considered in \ref\ggw{S.~J.~Gates, S.~Gukov and
E.~Witten, ``Two-Dimensional Supergravity Theories From
Calabi-Yau Four Folds'', \np {584} {2000} {109}, hep-th/0005120.},
\ref\hlm{M.~Haack, J.~Louis and M.~Marquart, ``Type IIA and
Heterotic String Vacua in $D=2$'', hep-th/0011075.}, \ref\GH{
S.~Gukov and M.~Haack, ``IIA String Theory on Calabi-Yau
Four-folds with Background Fluxes'', \np {639} {2002}  {95},
hep-th/0203267.}.} were derived in {\bb}. These constraints tell
us, that the internal component of the four-form $F$ of ${\cal
M}$-theory is of type $(2,2)$ and satisfies the primitivity
condition \eqn\aii{F\wedge J=0.} It was shown in
\ref\gvw{S.~Gukov, C.~Vafa and E.~Witten, ``CFT's from Calabi-Yau
Four-Folds'', \np {484} {2000} {69}, hep-th/9906070.}, that these
conditions can be derived from two superpotentials, that describe
the vacuum solutions in three dimensions \foot{In a nice paper by
De Wolfe and Giddings \ref\GDW{O.~De~Wolfe and
S.~B.~Giddings``Scales and Hierarchies in warped Compactifications
and Brane Worlds '', \np {584} {2000} {109}, hep-th/0208123.} the
effect of the warp factor on the superpotentials has been
computed.}
\eqn\aii{W=\int_{Y_4} F \wedge \Omega,}
and
\eqn\aiii{{\hat W}={1\over 4} \int_{Y_4}F\wedge J \wedge J.}
Here $\Omega$ describes the holomorphic four-form of the
Calabi-Yau four-fold $Y_4$, while $J$ describes the K\"ahler
form. Strictly speaking {\aiii} is not a
superpotential, as it is not a holomorphic function of the
K\"ahler moduli.
Nevertheless, we will be using this terminology, as it has become
standard in the literature. The above superpotentials and the
corresponding scalar potential were derived from a Kaluza-Klein
compactification of the ${\cal M}$-theory action in
\ref\hlt{M.~Haack and J.~Louis, ``${\cal M}$-Theory Compactified
on Calabi-Yau Four-folds with Background Flux'', \pl {507} {2001}
{296}, hep-th/0103068.}. In supergravity theories with four
supercharges the conditions for unbroken supersymmetry for
compactifications to three-dimensional Minkowski space are {\gvw}
\eqn\aiv{W=D_{\alpha}W=0\ ,}
and
\eqn\av{{\hat W}={D}_A{\hat W}=0\ .}
Here $\a=1,\dots,h^{13}$ describe the deformations of the complex structure
and $A=1,\dots,h^{11}$ parametrize the deformations of the K\"ahler
structure.
In these formulas the derivatives are defined as
$D_{\alpha}W={\partial}_{\alpha}W+({\partial}_{\alpha}K)W$ and
$D_{A}\hat W={\partial}_A {\hat W}-{1\over 2} ({\partial}_AK){\hat W}$,
where
\eqn\avi{K=-\log \int_{Y_4}\Omega \wedge { \bar \Omega}+\log {\cal V},}
is the three-dimensional K\"ahler potential.
The constraints for a supersymmetric three-dimensional vacuum
found in {\bb}, then easily follow from the above conditions
{\gvw}. Let us briefly go through the argument.
First, $W=0$ implies $F_{4,0}=F_{0,4}=0$. Second,
the identity
$D_{\alpha}W=\int {\Phi}_{\a} \wedge F $, where
${\Phi}_{\a}$ is a basis of $(3,1)$-forms, gives together with
$D_{\alpha}W=0$ the condition $F_{1,3}=F_{3,1}=0$.
Third, the condition $D_A{\hat W}=0$ implies together with
${\hat W}=0$, that $F$ is primitive.
However, Calabi-Yau four-folds are not the whole story, as the resulting
theory is three-dimensional and has an ${\cal N}=2$ supersymmetry.
Compactifications on other
Riemannian manifolds of exceptional holonomy are of special interest, as
they
allow
us to obtain theories with less supersymmetry and in a different number of
space-time dimensions. Recall, that there is a close connection between the
theory of
Riemannian manifolds with reduced holonomy and the theory of calibrated
geometry
\ref\HL{R.~Harvey and H.~B.~Lawson,
``Calibrated Geometries'', Acta Mat.{\bf 148} (1982) 47.}.
Calibrated geometry is the theory, which studies calibrated submanifolds, a
special
kind of minimal submanifolds of a Riemannian manifold, which are defined
using a closed form called the calibration. Riemannian manifolds with
reduced holonomy
usually come equipped with one or more natural calibrations. Based on this
close
relation to calibrated geometry and generalizing the result for the
superpotential
found in {\gvw}, Gukov
made a conjecture about the form of
the superpotential appearing in string theory compactifications
with
non-vanishing Ramond-Ramond fluxes
on a manifold $X$ of reduced holonomy
\ref\GU{S.~Gukov,
``Solitons, Superpotentials and Calibrations'', \np {574} {2000} {169},
hep-th/9911011.}
\eqn\avii{W=\sum \int_X({\rm calibrations})\wedge ({\rm Fluxes}).}
In this formula we sum over all possible combinations of
fluxes and
calibrations.
This conjecture has been checked by computing the scalar potential
from a Kaluza-Klein reduction of the action for a certain type of
theories. This in turn, determines the superpotential.
For the Type IIB theory these potentials have been computed in
\ref\TV{T.R.~Taylor and C.~Vafa,
``R-R Flux on Calabi-Yau and Partial Supersymmetry Breaking'',
\pl {474} {2000} {130}, hep-th/9912152.} and
\ref\gkp{S.~B.~Giddings, S.~Kachru and J.~Polchinski,
``Hierarchies from Fluxes in String Compactifications'',
hep-th/0105097.}. The superpotentials for Type IIA compactifications
on Calabi-Yau four-folds were derived
in {\ggw}, {\hlm}, {\GH}, while the scalar potential for ${\cal M}$-theory
on
$G_2$-holonomy manifolds has been computed in
\ref\bw{C.~Beasley and E.~Witten,
``A Note on Fluxes and Superpotentials in ${\cal M}$-theory
Compactifications
on Manifolds of $G_2$ Holonomy'', hep-th/0203061.}. Our goal will be to
compute the superpotential for two different theories.
Rather than computing the scalar potential
and from there obtain the superpotential, we shall compute the
superpotential directly by a Kaluza-Klein compactification of the gravitino
supersymmetry transformation. We shall illustrate the idea in section 2.
In section 3 we
will compute the superpotential for the heterotic
string compactified on a Calabi-Yau three-fold. It is well known, that for a
conventional compactification of the heterotic string on a Calabi-Yau
three-fold, i.e. without taking warp factors into account, turning on an
expectation
value
for the heterotic three-form will induce a superpotential, which breaks
supersymmetry without generating a cosmological constant
\ref\drsw{M.~Dine, R.~Rohm, N.~Seiberg and E.~Witten,
``Gluino Condensation in Superstring Models'',
\pl {156} {1985} {55}.}. In the context of Gukov's conjecture {\GU},
it was argued in
\ref\begu{K.~Behrndt and S.~Gukov, ``Domain Walls and Superpotentials from
M-Theory on Calabi-Yau Three-Folds'',
\np {580} {2000} {225}, hep-th/0001082.}, that this
superpotential can be written as in {\avii}, generalizing
the original proposal {\GU} to fluxes of Neveu-Schwarz type\foot{See also
{\drsw}
for an earlier discussion of the superpotential.}.
We shall check this
conjecture by computing the superpotential explicitly from a Kaluza-Klein
reduction
of the gravitino supersymmetry transformation.
In section 4 we shall apply a similar approach to compute the
superpotential for ${\cal M}$-theory compactifications on a Spin(7) holonomy
manifold.
In the appendix we will review some relevant aspects of Spin(7) holonomy
manifolds.

\newsec{Gauge Invariant Supergravity Lagrangian}
In order to derive the superpotential for the four-dimensional
heterotic string, it is easiest to compactify
the gravitino supersymmetry transformation law.
Recall, that the most general gauge invariant ${\cal N}=1$,
$D=4$ supergravity action can be described in terms of three functions (see
e.g.
\ref\ggrs{S.~J.~Gates, M.~T.~Grisaru, M.~Rocek and W.~Siegel,
``Superspace or One Thousand and One Lessons in Supersymmetry'',
Frontiers in Physics.},
\ref\wb{ J.~Wess and J.~Bagger, ``Supersymmetry and Supergravity'',
Princeton
Series in Physics.}). These are the superpotential $W$, the K\"ahler
potential $K$ and
a holomorphic function $H_{ab}$, which plays the role of the gauge coupling.
In the following we will take $H_{ab}={\d}_{ab}$.
The theory is formulated in terms of massless chiral multiplets, containing
a
complex scalar
$\phi$ and a Weyl spinor $\psi$ and  massless vector multiplets, containing
the
field $A^a_{\m}$ with field strength
$F^a_{\m\n}$ and a Weyl spinor ${\l}^a$. We shall be adding a real auxiliary
field
$D^a$ to the vector multiplets.
The bosonic part of the Lagrangian takes the following form
\foot{We will be following the conventions of {\wb}.}
\eqn\aixbb{{\cal L}=-{1\over 2}R-K_{{\bar
i}j}D_{\m}{\phi}^{i*}D^{\m}{\phi}^j
-{1\over 4} F^a_{\m\n}F^{a\m\n}-V(\phi,{\phi}^*).}
Here $V(\phi,\phi^*)$ describes the scalar potential given by
\eqn\aix{ V(\phi,\phi^*)=exp{(K)(K^{{\bar i}j}W^*_iW_j-3W^*W})
+{1\over 2} D^aD_a.}
In this formula $K^{{\bar i}j}$ is the inverse matrix to $K_{{\bar i}j}$,
\eqn\bi{ K_{\bar i j}={\partial^2 K(\phi, \phi^*)\over \partial
\phi^{i*}\partial \phi^j},}
and $W_i=\partial_iW+\partial_iK~W$.
The complete Lagrangian is invariant under ${\cal N}=1$ supersymmetry.
The relevant part of
the supersymmetry transformations takes the form
\eqn\ax{
\eqalign{
\delta {\l}^a & =F^a_{\m\n}~{\s^{\m\n}}\e -i D^a~{\e} , \cr
\d \psi_\m & = 2\nabla_{\m}\e +i~e^{K/2}~{\g}_{\m}{\e^*}W.\cr}}
Here ${\l}^a$ and ${\psi}_{\m}$ are positive chirality Weyl
spinors, describing the gluino and gravitino respectively, $\e$ is
a four-dimensional Weyl spinor of positive chirality, while
${\e^*}$ is the complex conjugate spinor with negative chirality.
If the space-time is flat, the complete supersymmetry
transformations tell us that supersymmetry demands (see {\wb})
\eqn\bixa{W_i=D^a=W=0.}
In the next section we will use the above supersymmetry
transformations to determine the superpotential and $D$-term for the
heterotic
string compactified on a Calabi-Yau three-fold.

\newsec{Superpotential for the Heterotic String on a Calabi-Yau Three-fold }
It has been known for a long time, that gluino
condensation triggers spontaneous supersymmetry
breaking in the heterotic string compactified on a Calabi-Yau three-fold
$Y_3$ (with no warp factors)
without producing a vacuum energy
{\drsw}.
In this process the Neveu-Schwarz three-form $H$ of the heterotic string
acquires
a vacuum expectation value proportional to the holomorphic three-form
${\Omega}$
of the Calabi-Yau three-fold. It was shown in {\drsw}, that this generates a
superpotential, which will break supersymmetry completely.
In a more recent context\foot{For an earlier discussion of the form of the
superpotential see {\drsw}.}
it was argued in
{\begu}, that the superpotential which is
induced by such a non-vanishing $H$-field
\eqn\axba{W=\int_{Y_3}H\wedge \Omega,}
extends the conjecture {\avii} to superpotentials with non-vanishing
fluxes
of Neveu-Schwarz type.
The argument, which motivated the above formula, was based on the
identification
of $BPS$ domain walls with branes wrapped over supersymmetric cycles.
More concretely, the $BPS$ domain wall of the ${\cal N}=1$, $D=4$ theory
originates from the heterotic
five-brane wrapping a special Lagrangian submanifold of $Y_3$. This is
because the five-brane is a source for the Neveu-Schwarz three-form field
strength $H$.
Here we would like to compute the form of the superpotential
and the form of the D-term appearing in {\ax} in this particular model by a
direct Kaluza-Klein reduction of the
gravitino and gluino supersymmetry transformation respectively. We then
would like to compare the result with formula {\axba}.
Recall that the ten-dimensional ${\cal N}=1$ supergravity multiplet contains
a metric
$g_{MN}$, a spin-${3\over 2}$ field ${\Psi}_M$, a two-form potential
$B_{MN}$,
a spin-${1\over 2}$ field $\l$ and a scalar field $\phi$. The
super-Yang-Mills
multiplet contains the Yang-Mills field $F^a_{MN}$ and a spin-${1\over 2}$
field ${\chi}^a$, the so called gluino.
The relevant part of the ${\cal N}=1$ supersymmetry transformations
in the ten-dimensonal string frame take the form
\eqn\axiiib{
\eqalign{\d \Psi_{\m} & = \nabla_{\m}\eta +{1 \over 48}
({\gamma}_{\mu}{\gamma}_5 ~{\otimes}~ \gamma^{abc} H_{abc})~ \eta, \cr
\delta {\chi}^{\a} & =-{1\over 4} F^{\a}_{ab} {\gamma}^{ab} \eta. \cr}
}
Here $\m$ describes the coordinates of the four-dimensional
Minkowski space, and $a,b, \dots$ describe the six-dimensional
internal indices, while ${\a}$ describes the gauge index.

We consider a Majorana representation for ten-dimensional Dirac
matrices with $\Gamma_M$ real and hermitian, apart from $\Gamma_0$
which is real and antihermitian. The matrices $\Gamma_M$ can be
represented as tensor products of $\gamma_{\mu}$, the matrices
of the external space, with $\gamma_m$, the matrices
of the internal space
\eqn\pingoi{ \eqalign{ \Gamma_{\mu} & = \gamma_{\mu} \otimes 1,
\cr \Gamma_m & = \gamma_5 \otimes \gamma_m, \cr } }
with
\eqn\pingoii{ \eqalign{ \gamma_5 = {i \over 4!} \,
\epsilon_{\m\n\r\s} \, \gamma^{\m\n\r\s}. \cr } }
We can also introduce the matrix
\eqn\pingoiii{ \eqalign{ \gamma = {i \over 6!}\sqrt{g_{(6)}} \,
\epsilon_{mnpqrs} \, \gamma^{mnpqrs}, \cr } }
which determines the chirality in the internal space. Here
$g_{(6)}$ represents the determinant of the internal metric. Thus
$\gamma_{\mu}$ are real and hermitian apart from $\gamma_0$ which
is real and antihermitian and $\gamma_m$ are imaginary and
hermitian as are $\gamma_5$ and $\gamma$. The relation between
$\Gamma$, the matrix which determines the chirality in
ten-dimensions, $\gamma_5$ and $\gamma$ is
\eqn\pingoiv{ \eqalign{ \Gamma = - \, \gamma_5 \otimes \gamma. \cr
} }
Consider $\eta$ a ten-dimensional Majorana-Weyl spinor of positive
chirality. In order to compactify transformations {\axiiib} to
four dimensions, we decompose this ten-dimensional spinor in terms
of the covariantly constant spinors of the internal manifold:
\eqn\eiviv{\eta=\e^* ~{\otimes}~\xi_++\e ~{\otimes}~\xi_{-},}
where $\xi_+$ and $\xi_- = (\xi_+)^*$ are six-dimensional Weyl
spinors with positive and negative chirality respectively and $\e$ is
a four-dimensional Weyl spinor of positive chirality, whose
complex conjugate is $\e^*$. Similarly, we decompose the
ten-dimensional gravitino as:
\eqn\eivvi{\Psi_{\m}=\psi^*_{\m} ~{\otimes}~
{\xi_+}+\psi_{\m}~{\otimes}~{\xi_-},}
where $\psi_{\m}$ is a four-dimensional Weyl spinor of positive
chirality, that represents the four-dimensional gravitino.

In complex coordinates the gravitino supersymmetry transformation
takes the form
\eqn\eivvii{ \eqalign{ \d \Psi_{\m} & = \nabla_{\m}\eta +{1 \over
48} \bigl ({\gamma}_{\m}{\gamma}_5 ~{\otimes}~ ( \gamma_{mnp}
H^{mnp}+ \gamma_{\bar m\bar n\bar p} H^{\bar m\bar n\bar p})\bigr)
~ \eta \cr &+{1 \over 48}\bigl ({\gamma}_{\m}{\gamma}_5
~{\otimes}~ (\gamma_{mn{\bar p}} H^{mn{\bar p}}+\gamma_{m{\bar
n}{\bar p}}H^{m{\bar n}{\bar p}}) \bigr)~ \eta. \cr}}
To evaluate the resulting expressions we use the identities (see
e.g. \ref\bbs{K.~Becker, M.~Becker and A.~Strominger,
``Five-Branes, Membranes and Nonperturbative String Theory'', \np
{456} {1995} {130}, hep-th/9507158.} or
\ref\mmms{M.~Marino, R.~Minasian, G.~Moore and A.~Strominger,
``Nonlinear Instantons from Supersymmetric p-Branes'', JHEP 0001:005,2000,
hep-th/9911206.})
\eqn\eivv{ \eqalign{\gamma_{\bar{m}}\xi_+&=0, \cr}}
\eqn\eivviii{ \eqalign{ \gamma_{ m n p} \xi_+ & = \|\xi_+\|^{-2}
~\Omega_{ m n p} ~\xi_-, \cr \gamma_{mn{\bar p}} \xi_+ & = 2i
~\gamma_{\lbrack m}J_{n \rbrack {\bar p}}~\xi_+ ,\cr
\gamma_{m{\bar n}{\bar p}} \xi_+ & = \gamma_{{\bar m}{\bar n}{\bar
p}} \xi_+ = 0. \cr}}
We now decompose our ten-dimensional spinors as in {\eiviv} and
{\eivvi} and make use of formulas ${\eivv}$ and ${\eivviii}$. Multiplying the resulting
expression from the left with $\xi_-^{\dagger}=\xi^T_+$, we obtain
the transformation:
\eqn\eivxab{{\d}{\psi}_{\m}={\nabla}_{\m}\e-{1 \over
48}{\g}_{\m}\e^* \, \|\xi_+\|^{-2} ~ H_{{\bar m}{\bar n}{\bar p}}
\, \Omega^{{\bar m}{\bar n}{\bar p}}.}
After integration over the internal manifold we obtain:
\eqn\eivxab{{\d}{\psi}_{\m}={\nabla}_{\m}\e+i{\g}_{\m}\e^* \,
\|\xi_+\|^{-2}\, e^{K_2} ~\int_{Y_3} H \wedge \Omega.}
where we have used that:
\eqn\eivxp{V ={1 \over 48}~\int_{Y_3} J \wedge J \wedge J={1 \over
64} \, e^{-K_2},}
with $V$ being the volume of the internal Calabi-Yau manifold. If
we choose
\eqn\eivxabo{\|\xi_+\|^{-2}=e^{K/2-K_2}}
and rescale the fields:
\eqn\eivxabp{\eqalign{\psi& \to {\psi \over 2}, \cr H& \to {H
\over 2}, \cr}}
then we obtain the four-dimensional supersymmetry transformation
for gravitino,
\eqn\eivxabs{{\d}{\psi}_{\m}=2{\nabla}_{\m}\e+i{\g}_{\m}\e^* \,
e^{K/2} ~\int_{Y_3} H \wedge \Omega.}
In the above formulas $K=K_1+K_2$ is the total K\"ahler potential, where $K_1$ is the 
K\"ahler potential for complex structure deformations
\eqn\eivix{K_1=-\log \left( i~\int_{Y_3}\O \wedge \bar{\O}
\right),}
and $K_2$ is the K\"ahler potential for the K\"ahler deformations
\eqn\eivx{K_2=-\log \left( {4 \over 3} ~\int_{Y_3} J \wedge J
\wedge J \right).}
Comparing this result with {\ax} we find the superpotential
\eqn\eivxiv{ W =\int_{Y_3} H \wedge \O,}
as promised.

Let us now consider the gluino supersymmetry transformations in
{\axiiib}. If we again decompose the gluino as in {\eivvi} and the
spinor $\eta$ as in {\eiviv}, we obtain after comparing with {\ax}
the form of the four-dimensional D-term up to a multiplicative
constant
\eqn\eab{D^a=F^a_{m\bar n}J^{m\bar n}.}
Here we have used
\eqn\eivxvi{J_{m\bar{n}}=-i \xi_+^{\dagger} \g_{m\bar{n}} \xi_+,}
while the expectation value for the other index contractions appearing in
the  four-dimensional
gluino supersymmetry transformation vanish.
As we have mentioned in the previous section, supersymmetry demands
$D^{(a)}=0$, which
in this case gives the, well known, Donaldson-Uhlenbeck-Yau equation
\eqn\eivxviii{J^{m\bar{n}} F_{m\bar{n}}^{a}=0.}
The fact that the Donaldson-Uhlenbeck-Yau equation originates from a D-term
constraint was first
discussed in
\ref\wittnew{E.~Witten, ``New Issues In Manifolds Of SU(3) Holonomy''
\np {268} {1986} {79}.}.
Furthermore, supersymmetry demands
\eqn\eivxix{W_i=0,}
where we are using again the notation
$W_i=\partial_iW+\partial_iK_1~W$, where $K_1$ is the K\"ahler
potential for complex structure deformations and is given in
{\eivix}. It is straightforward to evaluate this constraint to
obtain $W_i=\int_{Y_3} {\phi}_i\wedge  \Omega=0$, where ${\phi}_i$
is a complete set of $(2,1)$ forms \ref\caso{P.~Candelas and
X.~C.~de la Ossa,  ``Moduli Space of Calabi-Yau Manifolds'' \np
{355} {1991} {456}.}. This implies that $H$ is of type  $(0,3)$.
However in this case
\eqn\eivxx{W \not= 0,}
and we therefore see, that no supersymmetric solutions can be
found. It is expected, that  this situation changes, if we
consider instead a `warped' compactification of the heterotic
string {\str}, {\wsd}. The resulting background is in this case a
complex manifold with  non-vanishing torsion \foot{Manifolds with
non-vanishing torsion have also been discussed some time ago in
e.g. \ref\gates{S.~J.~Gates,``Superspace Formulation of New
Nonlinear Sigma Models'', \np {238}  {1984} {349}.}. As opposed
to the previous reference, the manifolds we shall be interested
in have  a torsion that is not closed.}. It is expected that
supersymmetric ground states can be found in this case. In
\ref\bbdg{K.~Becker, M.~Becker, K.~Dasgupta and P.~S.~Green, work
in progress.} we have already computed the form of this
superpotential and checked, that it takes the same form as
{\eivxiv}, but now the complete Chern-Simons terms of $H$ have to
be taken into account
\eqn\eivxxax{H=dB+{\omega}_L-{\omega}_{YM}.}
In the supergravity approximation, that we have been using, it is
only possible to take the Chern-Simons term of the gauge field
into account ${\omega}_{YM}$, as the Chern-Simons term
${\omega}_L$ coming from the spin connection is a higher order
effect. We shall report on more details elsewhere {\bbdg}. At the
same time, there is not much known about the mathematical
properties of these background manifolds, as many of the theorems
on complex manifolds do not generalize easily to the case of a
manifold with non-vanishing torsion.  Very generally, not many
examples of such manifolds with torsion are known. A few of them
have  been constructed in {\drs}. A rather interesting concrete
example of such a compactification of  the heterotic string on a
manifold with torsion has recently appeared \ref\bd{K.~Becker and
K.~Dasgupta, ``Heterotic Strings with Torsion'',
hep-th/0209077.}. Here it was shown, that  the supersymmetry
constraints derived in {\str} are satisfied for this particular
background  manifold. Very generally, it would be interesting to
understand the mathematical properties of these manifolds with
torsion. Work in this direction is in progress {\bbdg}.

\newsec{The Superpotential for Spin(7) Holonomy Manifolds}
In this section we would like to consider warped compactifications
of ${\cal M}$-theory on a Spin(7) holonomy manifold $X$. The
resulting action has an ${\cal N}=1$ supersymmetry in three
dimensions. These theories are closely related to four-dimensional
counterparts with completely broken supersymmetry. This is because
the dimensional reduction of the minimal ${\cal N}=1$, $D=4$
supersymmetric theory to three dimensions would lead to an ${\cal
N}=2$ supersymmetric theory. Models with ${\cal N}=1$
supersymmetry in three dimensions are interesting in connection to
the solution of the cosmological constant problem along the lines
proposed by Witten in \ref\witteno{E.~Witten, ``Is Supersymmetry
Really Broken? '' \ijmp {10} {1995} {1247}, hep-th/9409111.} and
\ref\witten{E.~Witten, ``Strong Coupling and the Cosmological
Constant'', \mpl {10} {1995} {2153}, hep-th/9506101.}. The basic
idea of this proposal is, that in three dimensions supersymmetry
can ensure the vanishing of the cosmological constant, without
implying the unwanted Bose-Fermi degeneracy. However, this
mechanism does not explain, why the cosmological constant of our
four-dimensional world is so small, unless there is a duality
between a three-dimensional supersymmetric theory and a
four-dimensional non-supersymmetric theory of the type, that we
are discussing. So, ${\cal M}$-theory compactifications on Spin(7)
holonomy manifolds allow us to address the cosmological constant
problem from the three-dimensional perspective.

The mathematical aspects of Spin(7) holonomy manifolds have extensively been
studied
in the literature
(see e.g.
\ref\jb{D.~D.~Joyce, ``Compact Manifolds with Special Holonomy'', Oxford
University Press.}, where examples of compact manifolds have been
discussed).
Recently, there have also been constructed examples of such manifolds, which
are not compact
\ref\cglp{M.~Cvetic, G.~W.~Gibbons, H.~Lu and C.~N.~Pope,``New Complete
Noncompact
Spin(7) Manifolds'',
\np {620} {2002} {29}, hep-th/0103155.} and
\ref\Gusp{S.~Gukov and J.~Sparks,``M Theory on Spin(7) Manifolds I'',
\np {625} {2002} {3}, hep-th/0109025.}.

Much less is known about the form of the most general
three-dimensional action with ${\cal N}=1$ supersymmetry,
describing the coupling of gauge fields to supergravity. Some
aspects, such as the field content of compactifications of ${\cal
M}$-theory on Spin(7) holonomy manifolds have been studied in
\ref\pato{G.~Papadopoulos and P.~K.~Townsend, ``Compactifications
of D=11 Supergravity on Spaces of Exceptional Holonomy'', \pl
{357} {1995} {300}, hep-th/9506150.}. However, the complete form
of the low energy effective action is not known at this point and
work in this direction is in progress \ref\bcgm{M.~Becker,
D.~Constantin, S.~J.~Gates, W.~Merrill and J.~Phillips, work in
progress.}. Very generally, it is known that the manifold of
scalars is in this case Riemannian instead of K\"ahler. It is
expected, that the metric on this Riemannian manifold can be
determined in terms of a potential function \foot{We thank
G.~Papadopoulos for discussions on this.}
\eqn\axibx{{\cal K}=-\log\int_{X} \Omega\wedge \Omega,}
where $\Omega$ describes the Cayley calibration of the Spin(7) holonomy
manifold.
This is a closed and self-dual four-form $\Omega=*\Omega$.
Furthermore, we expect that
the three-dimensional
action includes a superpotential, whose concrete form has
been conjectured in {\GU}
\eqn\axi{W=\int_{X} \Omega \wedge F.}
The constraints imposed by supersymmetry on these
compactifications were derived in \ref\katy{K.~Becker, ``A Note on
Compactifications on Spin(7)-Holonomy Manifolds'', JHEP 0105:003,
2001, hep-th/0011114.} and \ref\hata{S.~W.~Hawking and
M.~M.~Taylor-Robinson, ``Bulk Charges in Eleven-Dimensions'',
\physrev {58} {1998} {025006}, hep-th/9711042.}. In
\ref\ago{B.~Acharya, X.~de la Ossa and S.~Gukov, ``G Flux,
Supersymmetry and Spin(7) Manifolds'', hep-th/0201227.} it was
shown, that these constraints can be derived from the
superpotential {\axi}. Here we will check more  directly, that the
superpotential is given by {\axi}, by performing a Kaluza-Klein
reduction of the gravitino supersymmetry transformation law, along
the lines of the previous section. The bosonic part for
eleven-dimensional supergravity lagrangian contains a three-form
$C$ with field strength $F$ and the dual seven-form $*F$, as well
as the space-time metric $g_{MN}$
\eqn\axii{
{\cal L}= { 1\over 2} \int d^{11} x \sqrt{-g}
\left( R -{1 \over 2} F\wedge *F
-{1 \over 6} C\wedge F\wedge F\right).
}
Here we have set the gravitational constant equal to one.
The supersymmetry transformation of the gravitino $\Psi_M$ takes
the form
\eqn\axiii{
\eqalign{
\d \Psi_M & = \nabla_M\eta -{1 \over 288}
({\G_M}^{PQRS}-8 \d_M^P \Gamma^{QRS} )F_{PQRS} \eta,\cr}
}
where capital letters denote eleven-dimensional indices and $\eta$ is an
eleven-dimensional anticommuting Majorana spinor.
In order to compactify this theory on a Spin(7) holonomy manifold, we
will make the following ansatz for the metric
\eqn\axiv{
g_{MN}(x,y)=
\Delta^{-1}(y) \pmatrix{
 g_{\m \n}(x) & 0 \cr
0 &  g_{mn}(y)\cr
},
}
where $x$ are the coordinates of the external space labelled by the indices
$\mu,\nu,\dots$ and $y$
are the coordinates of the internal manifold labelled by $m,n,\dots$, while
$\Delta=\Delta(y)$ is
the warp factor. The eleven-dimensional spinor $\eta$ is decomposed as
\eqn\axv{
\eta=\epsilon \otimes \xi,
}
where $\epsilon$ is a three-dimensional anticommuting Majorana spinor and
$\xi$ is an eight-dimensional Majorana-Weyl spinor.
Furthermore, we will
make the following decomposition of the gamma matrices
\eqn\axvi{
\eqalign{
\Gamma_{\mu}=\gamma_{\mu} \otimes \gamma_9,  \cr
\Gamma_m= 1 \otimes \gamma_m , \cr
}
}
where $\gamma_{\mu}$ and $\gamma_m$ are the gamma matrices of the
external and internal space respectively. We choose the matrices
$\gamma_m$ to be real and antisymmetric. $\gamma_9$ is the eight-dimensional
chirality operator, which anti-commutes with all the $\gamma_m$'s.
In compactifications with maximally symmetric three-dimensional
space-time the non-vanishing components of the four-form field strength
$F$ are
\eqn\axvii{
\eqalign{&  F_{mnpq}\quad {\rm arbitrary} \cr
& F_{\mu \nu \rho m} =\epsilon_{\m\n\r} f_m , \cr}
}
where $\epsilon_{\m\n\r}$ is the Levi-Civita
tensor of the three-dimensional external space.
The external component of the gravitino supersymmetry transformation
is then given by the following expression {\katy}
\eqn\axviii{
\eqalign{
\delta \Psi_{\mu} =\nabla_{\mu} \eta & -{1 \over 288} \Delta^{3/2}
( \gamma_{\mu} \otimes  \gamma^{mnpq}) F_{mnpq} \eta \cr
& + {1 \over 6} \Delta^{3/2} ( \gamma_\mu  \otimes \gamma^m )
f_m \eta\cr
 & +{1 \over 4} \partial_n ( \log \Delta)
( \gamma_{\mu} \otimes \gamma^n ) \eta,  \cr}
}
where we have used a positive chirality eigenstate
$\gamma_9 \xi=\xi$. Considering negative chirality spinors
corresponds to an eight-manifold with a reversed orientation
\ref\ip{C.~J.~Isham and C.~N.~Pope, ``Nowhere vanishing Spinors
and Topological Obstruction to the Equivalence of the
NSR and GS Superstrings'',
\cqg {5} {1988} {257}, C.~J.~Isham, C.~N.~Pope and N.~P.~Warner,
``Nowhere-vanishing Spinors and Triality Rotations in 8-Manifolds'',
\cqg {5} {1988} {1297}.}.
We can decompose the eleven-dimensional gravitino as
\eqn\axviib{{\Psi}_{\mu}={\psi}^{(3)}_{\mu} \otimes \xi,}
where ${\psi}^{(3)}_{\m}$ is the three-dimensional gravitino.
After inserting {\axv} and {\axviib} in {\axviii}, we multiply both sides of
this
equation
from the left with the
transposed spinor $\xi^T$.
To evaluate the resulting expression we notice, that on these
eight-manifolds
it is possible to construct different types of $p$-forms
in terms of the eight-dimensional spinor $\xi$ as
\eqn\aii{
\omega_{a_1 \dots a_p} ={\xi}^T
{\gamma_{a_1\dots a_p}}{\xi}.
}
Since ${\xi}$ is Majorana-Weyl, {\aii} is non-zero only for
$p=0,4$ or 8 (see
\ref\gpp{
G.~W.~Gibbons, D.~N.~Page and C.~N.~Pope,
``Einstein Metrics on $S^3$, $R^3$ and $R^4$ Bundles'',
\cmp {127} {529} {1990}.}).
By this argument we notice, that the expectation values of the last
two terms appearing
in ${\axviii}$ vanish, as they contain only one internal gamma matrix.
The $Spin(7)$ calibration is given by the closed
self-dual 4-form {\hata}, {\katy}
\eqn\aii{
{ \Omega}_{mnpq } ={\xi}^T {\gamma}_{mnpq } {\xi}.
}
We have to be a little more careful though, as the previous form is defined
in terms of the covariantly constant spinor $\tilde \xi={\Delta}^{1/4}
{\xi}$
and gamma matrices that are rescaled by the warp factor in {\katy}.
However, to the order we are working out the supersymmetry transformation,
the warp factor can be taken to be constant.
We therefore obtain
from {\axviii}
\eqn\bxi{
\d \psi^{(3)}_\m = \nabla_{\m}\epsilon +{\gamma}_{\m} {\epsilon}
~\int_{X}F\wedge \Omega,}
where we have again dropped a multiplicative constant in front of the second
term
on the right hand side. Luckily from
\ref\bg{M.~Brown and S.~J.~Gates,
``Superspace Bianchi Identities and the Supercovariant Derivative'',
\ap {122} {1979} {443}.} it is known,
that the gravitino supersymmetry transformation in three dimensions
contains a term of a similar form as in the four-dimensional
case {\ax} but now formulated in terms of the three-dimensional
Majorana spinor. We can then read off the form of the superpotential
\eqn\axi{W=\int_{X} F \wedge \Omega,}
which is what we wanted to show.

Using this superpotential, it was shown in {\ago}, that the
${\cal N}=1$ supersymmetric
vacua in three-dimensional Minkowski space found in {\katy}
can be derived from the equations
\eqn\axibi{W={W_i}=0.}
Here ${W_i}$ indicates the derivative of $W$ with respect to the
scalar fields, that come from the metric deformations of the
Spin(7) holonomy manifold. From now on we will restrict our
analysis to compact Spin(7) holonomy manifolds, even though the
analysis of {\katy} was not restricted to that case. First we
notice, that if $X$ is an eight-manifold with Spin(7) holonomy,
then the internal component of the four-form $F$ can, in general,
belong to the following cohomologies
\eqn\axibix{H^4(X,\IR)=H_{1^+}^4(X,\IR)~
{\oplus}~ H_{27^+}^4(X,\IR)~{\oplus}~
H_{35^-}^4(X,\IR).}
The label $``\pm''$ indicates self-dual and anti-self-dual four-forms
respectively and the  subindex
indicates the representation. The Cayley calibration $\Omega$ belongs to the
cohomology
$H_{1^+}^4(X,\IR)$. Getting back to the equation {\axibi} we notice, that
the condition
$W=0$
implies $F_{1^+}=0$, which is equivalent to equation (24) of {\katy}.
According to {\ago}  and
{\jb}, $W_i$ generates the  $H^4_{35^-}(X,\IR)$ cohomology, so that $W_i=0$
implies  $F_{35^-}=0$.
This is equation (21) of {\katy}. There are no more constraints on the
fluxes for a compact
manifold, as in this case the cohomology group $H^4_{7^+}(X,\IR)$, vanishes.
To summarize,
supersymmetry demands that the flux on a compact Spin(7) holonomy manifold
takes the form
\eqn\axibix{F=F_{27^+}.}
We can now extend the arguments of {\katy} and {\ago},
to check, if it is possible to find
non-supersymmetric vacua with a vanishing three-dimensional cosmological
constant
as in {\gkp}, {\hlt},
\ref\bbt{M.~Becker and K.~Becker, ``Supersymmetry Breaking, ${\cal
M}$-Theory
and Fluxes, JHEP0107:038 (2001), hep-th/0107044.},
\ref\bbhl{K.~Becker, M.~Becker, M.~Haack and J.~Louis,
``Supersymmetry Breaking and Alpha-Prime Corrections to Flux Induced
Potentials'',
JHEP0206:060 (2002), hep-th/0204254.} and {\GH}.
A not supersymmetric solution to the equations of motion will satisfy
the condition $W_i=0$ but $W$ will be non-vanishing.
This means that for
compactifications on ${\cal M}$-theory on
Spin(7) holonomy manifolds the equations of
motion will be satisfied, if the internal component of the four-form field
strength
takes the form
\eqn\axibixx{F=F_{1^+}~{\oplus}~F_{27^+}.}
Supersymmetry demands the additional constraint $W=0$. As we had already
seen, this means  that the
first term on the right hand side of the above expression vanishes.
Therefore,
in this type of
compactifications it is possible to find a solution to the equations of
motion in  three-dimensional
Minkowski space, that breaks supersymmetry, by turning on the form
$F_{1^+}$, without  generating a
cosmological constant. Such an interesting scenario with a vanishing
cosmological constant  and
broken supersymmetry has already appeared in a number of different contexts
{\gkp}, {\hlt},  {\bbt},
{\bbhl} and {\GH}. However, in contrast to the superpotentials appearing in
the previous
references, it is expected that the superpotential {\axi} receives
perturbative and
non-perturbative quantum corrections. For an analysis of some aspects of
these corrections  see
{\ago}. This completes our discussion about ${\cal M}$-theory
compactifications on Spin(7)  holonomy
manifolds.

\newsec{Conclusions and Outlook}
In this paper we have checked the
conjecture
{\avii} made in {\GU} regarding the form of the superpotential, which is
induced when  non-trivial fluxes are turned on in different types of string
theory and ${\cal M}$-theory
compactifications. We do so, by performing a Kaluza-Klein reduction of the
gravitino
supersymmetry transformation.
As it is well known from {\drsw}, a compactification of the heterotic string
on a Calabi-Yau three-fold leads to a
superpotential, which breaks supersymmetry completely. We have checked, that
this  superpotential can be written in the form {\eivxiv}, which extends the
conjecture made
in {\GU} to fluxes of Neveu-Schwarz type {\begu}.

It is interesting, that the superpotential for warped compactifications of
the
heterotic string is given by {\eivxiv}, where the heterotic $H$ field now
includes the
Chern-Simons terms {\bbdg}.
For these compactifications, the internal manifold is then, in general,
non-K\"ahler and
has a non-vanishing torsion. Many theorems on complex manifolds do not
easily
generalize to
manifolds with non-vanishing torsion. Therefore, these string theory
compactifications are
far from being well understood. These issues are presently under
investigation
and we will report on this in a near future {\bbdg}.

In the second part of the paper we have considered warped
compactifications of ${\cal M}$-theory on Spin (7) holonomy manifolds. These
compactifications have been
previously analyzed in {\katy}, {\hata} and {\ago}.
Not much is known about the form of the low energy effective action of the
resulting three-dimensional ${\cal N}=1$ theory
and work in this direction is in progress {\bcgm}.
Luckily it is known, that the gravitino supersymmetry transformation
contains a
similar term as in the four-dimensional case {\bg}, so that we could
verify the conjecture of {\GU} by a direct calculation of the
superpotential.
Using this superpotential, we have shown the existence of solutions to
the three-dimensional equations of motion, which break supersymmetry and
have a vanishing
three-dimensional cosmological constant. Such an interesting scenario has
recently
appeared many times in the literature.

Contrary to the superpotential appearing in compactifications of ${\cal
M}$-theory
on Calabi-Yau four-folds, it is known that this ${\cal N}=1$ superpotential
receives  perturbative and non-perturbative quantum corrections {\ago}. It
would be nice
to compute these corrections along the lines of {\bbhl}.

Finally, it would be interesting to derive the constraints
{\axibixx} from the direct analysis of the equations of motion,
as in {\bbt}. This might be a little bit more complicated as
Spin(7) holonomy manifolds are, in general non-K\"ahler.

\vskip 1cm

\noindent {\bf Note Added}

The superpotential for compactifications of the heterotic string
on non-K\"ahler complex six-dimensional manifolds has now been
computed in \ref\bbhl{K.~Becker, M.~Becker, K.~Dasgupta and
P.~S.~Green ``Compactifications of Heterotic Theory on
Non-K\"ahler Complex Manifolds: I'', hep-th/0301161.}.

\vskip 1cm

\noindent {\bf Acknowledgement}

\noindent

We thank K.~Dasgupta, S.~J.~Gates, P.~S.~Green, S.~Gukov,M.~Haack,
W.~D.~Linch III, G.~Papadopoulos, J.~Phillips, E.~Witten and
especially K.~Becker for useful discussions. This work was
supported by NSF under grant PHY-01-5-23911 and an Alfred Sloan
Fellowship.

\vskip 1cm

{\noindent} {\bf Appendix I. Review of Spin(7) Holonomy Manifolds}

This appendix contains a brief review of some of the relevant aspects
of Spin(7)
holonomy manifolds. A very nice and complete discussion can be found in
{\jb}.
On an Riemannian manifold $X$ of dimension $n$, the spin connection
${\omega}$ is, in general,
an $SO(n)$ gauge field. If we parallel transport a spinor $\psi$ around a
closed path $\gamma$,
the spinor comes back as $U\psi$, where $U=Pexp\int_{\gamma}{\omega}\cdot
dx$ is the path
ordered exponential of $\omega$ around the curve $\gamma$.

A compactification of ${\cal M}$-theory (or string theory) on $X$ preserves
some amount of
supersymmetry, if $X$ admits one (or more) covariantly constant spinors.
Such spinors return upon
parallel transport to its original value, i.e. they satisfy $U\psi=\psi$.
The holonomy of the
manifold is then a subgroup of $SO(n)$. A Spin(7) holonomy manifold is an
eight-dimensional
manifold, for which one such spinor exists. Therefore, if we compactify
${\cal M}$-theory on these
manifolds we obtain an ${\cal N}=1$ theory in three dimensions. Spin(7) is a
subgroup of
$GL(8,\IR)$ defined as follows. Introduce on ${\IR}^8$ the coordinates
$(x_1,\dots,x_8)$ and the
four-form $dx_{ijkl}=dx_i\wedge dx_j\wedge dx_k\wedge dx_l$. We can define a
four-form  $\Omega$ on
${\IR}^8$ by
\eqn\ai{
\eqalign{\Omega&
=dx_{1234}+dx_{1256}+dx_{1278}+dx_{1357}-dx_{1368}\cr
&-dx_{1458}-dx_{1467}-dx_{2358}-dx_{2367}-dx_{2457}\cr
&+dx_{2468}+dx_{3456}+dx_{3478}+dx_{5678}.\cr}}
The subgroup of $GL(8,\IR)$ preserving $\Omega$ is the holonomy group
Spin(7). It is a  compact,
simply connected, semisimple, twenty-one-dimensional Lie group, which is
isomorphic to the  double
cover of $SO(7)$. The form $\Omega$ is self-dual, i.e. it satisfies
$\Omega=*\Omega$, where  $*$ is
the Hodge star of ${\IR}^8$. Many of the mathematical properties of Spin(7)
holonomy  manifolds are
discussed in detail in {\jb}. Let us here only mention that these manifolds
are Ricci flat  but are,
in general, not K\"ahler. The cohomology of a compact Spin(7) holonomy
manifold can be  decomposed
into the following representations of Spin(7)

\eqn\pppwe{ \eqalign{ H^0(X, \IR) & =  \IR \cr H^1(X, \IR) & =  0 \cr
H^2(X,\IR) &
=H^2_{{\bf21}}(X, \IR) \cr H^3 (X, \IR) & =  H^3_{{\bf 48}}(X, \IR) \cr H^4
(X, \IR) & =  H^4_{{\bf
1}^+} (X, \IR) \oplus H^4_{{\bf 27}^+} (X, \IR) \oplus H^4_{{\bf 35}^-} (X,
\IR) \cr
H^5 (X, \IR) &
= H^5_{{\bf 48}} (X, \IR) \cr H^6 (X, \IR) & = H^6_{{\bf 21}} (X, \IR)  \cr
H^7
(X, \IR) & = 0  \cr
H^8 (X,\IR) & = \IR \cr }}
The label $``\pm''$ indicates self-dual and anti-self-dual four-forms
respectively and the  subindex
indicates the representation. The Cayley calibration $\Omega$ belongs to the
cohomology
$H_{1^+}^4(X,\IR)$. In this decomposition one has to take into account, that
for a compact  Spin(7)
holonomy manifold $H^4_{7^+}(X,\IR)=0$.

\listrefs

\end